\documentclass[journal=jctc,manuscript=article]{achemso}
\usepackage{graphicx}
\usepackage{dcolumn}
\usepackage{bm}
\usepackage{rotating} 
\usepackage{epsfig}
\usepackage[version=3]{mhchem} 
\usepackage{cancel}
\usepackage{caption}
\usepackage{multirow}
\usepackage{subcaption}
\usepackage{qcircuit}
\usepackage{color, colortbl}
\usepackage{braket}
\usepackage[colorlinks=true, linkcolor=blue, citecolor=blue, urlcolor=blue]{hyperref}

\usepackage{amsthm}
\usepackage{amsmath}

\usepackage{thmtools}
\usepackage{thm-restate}
\usepackage{hhline}
\usepackage{orcidlink}
\SectionNumbersOn
 
\definecolor{Gray}{gray}{0.9}
\newcolumntype{a}{>{\columncolor{white}}c}
\newcolumntype{b}{>{\columncolor{Gray}}c}

\usepackage[normalem]{ulem}
\usepackage{color}

\title{Reducing the runtime of fault-tolerant quantum simulations in chemistry through symmetry-compressed double factorization}

\author{Dario Rocca\orcidlink{0000-0003-2122-6933}}
\email{dario.rocca@qcware.com}
\affiliation[QCW]{QC Ware Corporation, Palo Alto, California 94301, USA}

\author{Cristian L. Cortes\orcidlink{0000-0002-1163-2981}}
\affiliation[QCW]{QC Ware Corporation, Palo Alto, California 94301, USA}

\author{Jerome Gonthier\orcidlink{0000-0002-2933-4085}}
\affiliation[QCW]{QC Ware Corporation, Palo Alto, California 94301, USA}

\author{Pauline J. Ollitrault\orcidlink{0000-0003-1351-7546}}
\affiliation[QCW]{QC Ware Corporation, Palo Alto, California 94301, USA}

\author{Robert M.~Parrish\orcidlink{0000-0002-2406-4741}}
\affiliation[QCW]{QC Ware Corporation, Palo Alto, California 94301, USA}

\author{Gian-Luca Anselmetti\orcidlink{0000-0002-8073-3567}}
\affiliation[BI]{Quantum Lab, Boehringer Ingelheim, 55218 Ingelheim am Rhein, Germany}

\author{Matthias Degroote\orcidlink{0000-0002-8850-7708}}
\email{matthias.degroote@boehringer-ingelheim.com}
\affiliation[BI]{Quantum Lab, Boehringer Ingelheim, 55218 Ingelheim am Rhein, Germany}

\author{Nikolaj Moll\orcidlink{0000-0001-5645-4667}}
\affiliation[BI]{Quantum Lab, Boehringer Ingelheim, 55218 Ingelheim am Rhein, Germany}

\author{Raffaele Santagati\orcidlink{0000-0001-9645-0580}}
\affiliation[BI]{Quantum Lab, Boehringer Ingelheim, 55218 Ingelheim am Rhein, Germany}

\author{Michael Streif\orcidlink{0000-0002-7509-4748}}
\affiliation[BI]{Quantum Lab, Boehringer Ingelheim, 55218 Ingelheim am Rhein, Germany}

\begin{document}

\begin{abstract}
Quantum phase estimation based on qubitization  
is the state-of-the-art fault-tolerant quantum algorithm for computing ground-state energies in
chemical applications. In this context, the 1-norm of the Hamiltonian plays a fundamental role
in determining the total number of required iterations and also the 
overall computational cost. In this work, we introduce the symmetry-compressed double factorization (SCDF) approach, which combines a compressed double factorization of the Hamiltonian with the symmetry shift technique, significantly reducing the 1-norm value.
The effectiveness of this approach is demonstrated numerically by considering various benchmark
systems, including the FeMoco molecule, cytochrome P450, and 
hydrogen chains of different sizes. To compare the  efficiency of SCDF to other methods in absolute terms,
we estimate Toffoli gate requirements, which dominate the execution time on fault-tolerant
quantum computers.
For the systems considered here, SCDF leads to a sizeable reduction of the Toffoli gate count
in comparison to other variants of double factorization or even tensor hypercontraction, which is usually regarded as the most efficient approach for qubitization.
\end{abstract}

	\maketitle

\section{Introduction}

Quantum chemistry simulations hold a significant potential to advance many industry-relevant applications, including 
the development of new drugs \cite{heifetz2020quantum}, catalysts \cite{norskov2009towards}, and materials \cite{jain2013commentary}.
The simulation of chemical systems from first principles requires the solution
of the Schr\"odinger equation, a task particularly challenging
for classical approaches because of the exponential growth of the computational cost with system size. Quantum computing provides a promising solution to address this scalability issue, with
significant ongoing efforts focused on developing resource-efficient algorithms~\cite{Cao2019Quantum}.
Much of this work has been dedicated to approaches tailored for early-stage noisy quantum
devices, such as the variational quantum eigensolver (VQE) \cite{peruzzo2014variational, tilly2022variational}. Besides the challenges of working with noisy hardware,
optimizing the parameters in the VQE ansatz is non-trivial, and 
the number of required measurements grows rapidly with the system size.

The significant challenges to achieving a quantum advantage in near-term noisy devices 
motivate current efforts to transition towards fault-tolerant quantum computing (FTQC).
Early hardware demonstrations of error-corrected logical qubits have already been achieved \cite{google2023suppressing,bluvstein2023logical} 
and many companies, including IBM \cite{ibmroadmap}
and Google \cite{googleroadmap}, have announced roadmaps
to build fault-tolerant quantum computers in the next few years. At the same time, a parallel effort is underway to develop quantum algorithms that 
can efficiently exploit error-corrected qubits.

Quantum phase estimation (QPE) can be considered as the prototypical algorithm for
chemistry simulations within the FTQC framework \cite{abrams1999quantum, aspuru2005simulated}.
Within the standard formulation of QPE,
the calculation of the ground state energy of a given chemical system relies on the implementation of the Hamiltonian evolution operator $\mathcal{U}[\hat{H}]=e^{-i\hat{H}\tau}$
for some duration $\tau$;
this operator can be approximated in practice using the Trotter-Suzuki \cite{suzuki1993improved}
formula or Taylor series expansion \cite{berry2015simulating}. 
More recently, an alternative approach for QPE has been proposed based on the quantum walk operator 
$\mathcal{W}[\hat{H}]=e^{-i\arccos(\hat{H}/\lambda)}$ \cite{berry2018improved,poulin2018quantum,babbush2018encoding}.
Instead of directly returning the ground state energy, the algorithm outputs the arccosine of the ground state energy.
The advantage of this procedure is that the quantum circuit corresponding to
the quantum walk operator can be implemented exactly using qubitization \cite{low2019hamiltonian}.
The parameter $\lambda$ in the definition of $\mathcal{W}[\hat{H}]$ corresponds to the 1-norm of the Hamiltonian, and
its value is influenced by the specific representation of $\hat{H}$ and the strategy used to block encode it. This parameter plays a 
crucial role in the QPE efficiency, and its optimization is one of the 
main topics of this work.

Within the qubitization-based QPE approach, the total number of Toffoli gates scales as 
$\mathcal{O}\left( \frac{\lambda }{\epsilon}C_{\mathcal{W}[\hat{H}]} \right)$.
Here, $\epsilon$
represents the accuracy required for the ground state energy (typically, this should be within the chemical accuracy threshold of 1.6 mHa), and the ratio $\lambda/\epsilon$ determines the total number of iterations;
$C_{\mathcal{W}[\hat{H}]}$ is the Toffoli gate cost per iteration and depends on the specific 
approach used for implementing $\mathcal{W}[\hat{H}]$.
Implementing Toffoli gates or, similarly, T gates on quantum hardware requires a procedure known as magic state distillation \cite{bravyi2005universal,reichardt2005quantum,babbush2018encoding}. This process demands a considerably large number of qubits and takes significantly more time than other operations in the computation.
For this reason, a reduction in the overall Toffoli gate count
for a quantum algorithm is expected to lead to an equivalent reduction in the overall 
runtime.

While the 1-norm plays a fundamental role in determining the total number of iterations,
 $C_{\mathcal{W}[\hat{H}]}$ also has a significant contribution to the overall computational cost.
Specifically, the computational complexity
of realizing $\mathcal{W}[\hat{H}]$ depends on 
$\Gamma$, the amount of information needed
to specify the Hamiltonian, and the specific approach employed for the
quantum implementation. As discussed in Ref.~\citenum{lee2021even}, the combination of tensor
factorizations with
techniques such as unary iteration~\cite{babbush2018encoding} and optimized QROM assisted by ancillae~\cite{low2018trading,berry2019qubitization} leads to a $\mathcal{O}(\sqrt{\Gamma})$ Toffoli gate 
and logical qubits scaling. Further details on the origin of this square root dependence will be provided in Sec. \ref{sec:compdetails}. A summary of the 
computational complexity of 
state-of-the-art approaches for qubitization-based QPE 
is presented in Table \ref{tab:qubitizationcomplexity}.
Beyond a different cost in the $\mathcal{W}[\hat{H}]$
implementation, these approaches also involve 
different definitions of the 1-norm $\lambda$,
 whose scaling varies between $\mathcal{O}(N)$ and $\mathcal{O}(N^3)$, in which $N$ is the number of spatial orbitals the Hamiltonian is expressed in.

 \begin{table*}[tb!]
\centering
\begin{tabular}{l|l|l}
\hline  \hline
Approach   & Logical qubits   & Toffoli gates   \\ 
\hline \hline
Sparse method \cite{berry2019qubitization} & $\mathcal{O}(N+\sqrt{S})$ &  $\mathcal{O}((N+\sqrt{S})\lambda_V/\epsilon)$   \\
Single factorization \cite{berry2019qubitization} & $\mathcal{O}(N^{3/2})$  & $\mathcal{O}(N^{3/2}\lambda_{SF}/\epsilon)$ \\ 
Explicit double factorization\cite{von2021quantum} & $\mathcal{O}(N \sqrt{\Xi})$  & $\mathcal{O}(N \lambda_{\textrm{DF}} \sqrt{\Xi}/\epsilon)$ \\ 
Tensor hypercontraction\cite{lee2021even} & $\mathcal{O}(N)$ & $\mathcal{O}(N\lambda_{\textrm{THC}}/\epsilon)$ \\
\hline \hline
\end{tabular}
\caption{Asymptotic scaling of the computational resources required by different 
approaches used in qubitization-based quantum phase estimation. The definition of
the 1-norm $\lambda$ depends on the specific implementation. The ratio of $\lambda$ to
the required precision $\epsilon$ in the final result determines the total number of iterations. 
$N$ denotes the number
of spin orbitals, $S$ the sparsity of the Hamiltonian, and $\Xi$ the average rank of the
second factorization.}
\label{tab:qubitizationcomplexity}
\end{table*}

A straightforward implementation based on the electronic Hamiltonian in
second quantization involves $\mathcal{O}(N^4)$ terms. 
To improve over this complexity,
a sparse method was introduced
that truncates the components of the two-electron tensor 
according to a chosen threshold \cite{berry2019qubitization}.
The main limitation of this approach is that the
number of remaining terms $S$ in the Hamiltonian
cannot be systematically predicted, and in some cases, still
behaves as $\mathcal{O}(N^4)$. 
The single factorization (SF) approach applies an eigendecomposition to the two-electron tensor (see Eq.~\ref{firstfact} below)
and this effectively decreases the number of terms in the Hamiltonian to $\mathcal{O}(N^{3})$ \cite{berry2019qubitization}.
The explicit double factorization (XDF) approach introduces a second factorization on top of the SF (see Eq.~\ref{secondfact} below) \cite{motta2021low,kivlichan2018quantum,berry2019qubitization,huggins2021efficient}.
This reduces the Hamiltonian to $\mathcal{O}(N^{2}\Xi)$  pieces of information,
where $\Xi$ is the average rank of the second tensor factorization.
The numerical experiments for hydrogen chain considered in Sec.~\ref{sec:hchains}
show that $\Xi$ itself is characterized by a $\mathcal{O}(N)$ behavior.
As discussed in Sec.~\ref{sec:XDFandCDF}, an alternative approach known as compressed double factorization (CDF) builds the tensors in the factorization by optimizing a suitable cost function \cite{cohn2021quantum,oumarou2022accelerating, rubin2022compressing}. The new methodology presented in this paper will be based on a variant of the CDF approach.
The recent work of von Burg \emph{et al.}~\cite{von2021quantum} has introduced an efficient quantum algorithm to
implement the double-factorized Hamiltonian in the QPE framework by employing Givens rotations
and qubitization \cite{von2021quantum}. Compared to a straightforward qubitization of the double-factorized Hamiltonian, this formulation also benefits from significantly reducing the 1-norm.

The tensor hypercontraction (THC) approach decomposes the two-electron tensor in the Hamiltonian
as $g_{pqrs} \approx \sum_{\mu, \nu=1}^{N_{\textrm{THC}}} \chi_p^{(\mu)} \chi_q^{(\mu)} \zeta_{\mu \nu} \chi_p^{(\nu)} \chi_q^{(\nu)}$, where $\chi_p^{\mu}$ and $\zeta_{\mu \nu}$ denote the components of
the tensors used for this decomposition, $N_{\textrm{THC}}$ is the THC rank, and $p$, $q$, $r$, and $s$ are indices identifying the spatial orbitals \cite{hohenstein2012tensor,parrish2012tensor}. 
The tensors are obtained by minimizing a cost function
that determines the deviation of the decomposition from the exact two-electron tensor.
The application of this approach in the context of QPE was first proposed in Ref.~\citenum{lee2021even}. 
To effectively decrease the 1-norm, the $\chi$ tensors were used as basis set rotations,
applying them to redefine the representation of the corresponding creation and annihilation operators in the second-quantized Hamiltonian.
In practice, this amounts to reformulating the Hamiltonian in a larger non-orthogonal basis set
and new techniques were developed to block encode and qubitize it \cite{lee2021even}.
Beyond decreasing the 1-norm, the THC approach provides a very compact representation of the Hamiltonian, with $\Gamma=\mathcal{O}(N^2)$ and, correspondingly, an improved asymptotic
computational complexity. Since this method has systemically provided the most favorable
resource estimations for many examples of electronic Hamiltonians \cite{lee2021even, goings2022reliably}, it will serve as the main benchmark  
for the methodological developments proposed in this work.

This work is largely focused on the reduction of the 1-norm that 
has a strong impact on the number of iterations and, accordingly, on the 
total runtime of the QPE algorithm. 
Different approaches have been proposed in the literature to optimize the 
1-norm.
The XDF in the implementation of von Burg \emph{et al.}\cite{von2021quantum} and the THC \cite{lee2021even}
benefit themselves from formulations that significantly reduce the 1-norm as 
compared to a straightforward transformation of the electronic Hamiltonian
into Pauli words.
The optimization of the 1-norm for quantum simulations has been considered in 
previous work. Orbital transformations were proven to improve
the 1-norm values significantly \cite{koridon2021orbital}.  
In Ref.~\citenum{loaiza2022reducing}, several different approaches (including orbital transformation) were compared by considering small
molecules in the minimal STO-3G basis set; it was shown that double factorization coupled with a symmetry shift provides the best results in terms of 1-norm reduction and scaling with the
system size. This symmetry shift approach, described in detail in Sec. \ref{sec:symshift}, effectively decreases the
1-norm by subtracting a function of the number operator of electrons from the electronic Hamiltonian.
Since the number operator of electrons commutes with the Hamiltonian, the eigenvectors
of the Hamiltonian are not affected by this shift, and the correct ground state energy can be obtained
by applying a simple \emph{a posteriori} correction.

The new symmetry-compressed double factorization (SCDF) approach introduced here
exploits the symmetry shift idea but additionally optimizes the DF
tensor decomposition to further decrease the 1-norm.
Numerical demonstrations of this method include active space models of the FeMoco molecule and
cytochrome P450, and hydrogen chains with up to 80 atoms. For all of these systems,
SCDF, to the best of our knowledge, provides the smallest values of the 1-norm reported in the literature.
This leads to a Toffoli gate count and runtime that are sizeably reduced with respect to THC (for example by one half for FeMoco and P450). 
The SCDF factorization has the same structure of XDF and can be implemented using the techniques
proposed by von Burg \emph{et al.}~\cite{von2021quantum}.  
Accordingly, SCDF inherits an analogous 
computational complexity both in terms 
of Toffoli gate and logical qubit requirements (see Table \ref{tab:qubitizationcomplexity}) 
but with a 1-norm that scales more favorably with the number of orbitals compared to XDF (see Sec. \ref{sec:hchains}).
Despite a slightly worse asymptotic behavior than THC, the numerical applications 
considered in this work show that SCDF provides 
more systematic accuracy for ground state energies owing to a simpler numerical optimization scheme. This feature is crucial to
address systems of large size and to obtain reliable properties in the thermodynamic
limit.  

\section{Methodological approach}\label{sec:method}

\subsection{General double factorization framework}\label{sec:generalDF}

Within the second quantization formalism, the electronic Hamiltonian is expressed as
\begin{eqnarray}\label{finalsq}
    \hat{H} = E_{\textrm{nuc}} + \sum_{p,q=1}^N k_{pq} \hat{E}_{pq} +
    \frac{1}{2} \sum_{p,q,r,s=1}^N g_{pqrs} \hat{E}_{pq} \hat{E}_{rs},
\end{eqnarray}
where $p$, $q$, $r$, and $s$ are indices identifying the $N$ spatial orbitals, and the singlet spin-summed one-particle substitution operators are defined as $\hat{E}_{pq}\equiv 
\hat{a}^{\dag}_p \hat{a}_q + \hat{a}^{\dag}_{\bar{p}} \hat{a}_{\bar{q}}$. In this definition $\hat{a}^{\dag}$ and $\hat{a}$ denote creation and annihilation operators, respectively, and the bar on top of the orbital indexes indicates
a $\downarrow$ spin orbital.
 
In Eq. \ref{finalsq} the constant term $E_{\textrm{nuc}}$  
corresponds to the nuclear repulsion energy, 
\begin{eqnarray}
     g_{pqrs}=(pq|rs)= \iint d \mathbf{r}_1 \  d \mathbf{r}_2 \
    \phi_p(\mathbf{r}_1)  \phi_q(\mathbf{r}_1)
    \frac{1}{r_{12}}  
    \phi_r(\mathbf{r}_2)  \phi_s(\mathbf{r}_2)
\end{eqnarray}
is the two-electron tensor, and the modified one-electron tensor $k_{pq} = h_{pq} - \frac{1}{2} \sum_{r=1}^N g_{prrq}$ is defined in terms of the
one-electron integrals
\begin{eqnarray}
    h_{pq}=(p|h|q)= \int d \mathbf{r} \  
    \phi_p(\mathbf{r}) \left( -\frac{1}{2} \nabla^2 
    - \sum_I \frac{Z_I}{r_I} \right) \phi_q(\mathbf{r}),
\end{eqnarray}
that include the kinetic and electron-nucleus interaction energies. Without loss of generality for molecular systems, the spatial orbitals $\phi(\mathbf{r})$ have been chosen to be real.

The second quantized Hamiltonian can then be expressed
in a quantum computing amenable form 
by expanding it in terms of Pauli words using, for example, the Jordan-Wigner or Bravyi-Kitaev
transformations \cite{jordan1928paul,bravyi2002fermionic,seeley2012bravyi}.
These approaches lead to $\mathcal{O}(N^4)$ number of terms in the Hamiltonian. Despite the polynomial growth, this number of terms poses practical challenges for noisy near-term and fault-tolerant algorithms, and, in this context, the double factorization of the Hamiltonian can provide several advantages.

The main idea of double factorization consists in decomposing the two-electron tensor in 
the following way \cite{poulin2014trotter,motta2021low,kivlichan2018quantum,berry2019qubitization,matsuzawa2020jastrow,huggins2021efficient,cohn2021quantum}:
\begin{eqnarray}\label{DFmain}
    (pq|rs) \approx \sum_{t=1}^{N_{\textrm{DF}}} \sum_{k,l=1}^N U^{t}_{pk} U^{t}_{qk} V^t_{kl} U^{t}_{rl} U^t_{sl},
\end{eqnarray}
where the $\mathbf{U}^t$ tensors are orthonormal, namely
\begin{eqnarray}
    \sum_{k=1}^N U^t_{pk} U^{t}_{qk} = \delta_{pq},   \quad \quad
    \sum_{k=1}^N U^t_{kp} U^{t}_{kq} = \delta_{pq}, \label{ortho}
\end{eqnarray}
and the ``core'' tensors 
\begin{eqnarray}
    V^t_{kl}=V^t_{lk}
\end{eqnarray}
are symmetric for all $t$'s. 
The sum over $t$ in Eq.~\ref{DFmain} runs up to a maximum value $N_{\textrm{DF}}$ 
that depends on the specific approach used to build the
tensor factorization and crucially determines 
the trade-off between the accuracy and efficiency of the method. 
Depending on the specific DF implementation, the sums over $k$ and $l$ can also be limited to values $\Xi^{(t)}\leq N$.
The truncation of the tensors within the double factorization
approach will be further discussed below.

By inserting the factorized tensors of Eq.~\ref{DFmain} 
into Eq.~\ref{finalsq}, it is possible to reformulate the second quantized Hamiltonian
in terms of the operators 
$\tilde{a}_{kt}^{\dag}=\sum_{p=1}^N U^{t}_{pk} \hat{a}_{p}^{\dag}$ and
$\tilde{a}_{kt}=\sum_{q=1}^N U^{t}_{qk} \hat{a}_{q}$,
that create and annihilate electrons, respectively, in a new set of
rotated orbitals. In order to apply these operators, it is convenient to introduce the $\hat{G}_t$ operators, that rotate the quantum state in the new orbital basis and,
using the Thouless theorem \cite{thouless1960stability}, can be expressed as
\begin{eqnarray}
\hat{G}^{\dag}_t
= \exp \left( \sum_{p,q=1}^N [\log \mathbf{U}^t]_{pq} \hat{a}^{\dag}_p \hat{a}_q \right).
\end{eqnarray}
These rotations can be formulated in
terms of Givens rotations networks that can be efficiently implemented on
quantum hardware \cite{kivlichan2018quantum}.
Within the double-factorized formalism, the
Hamiltonian in second quantization can then be expressed as 
\begin{eqnarray}\label{intermediateSQ}
    \hat{H}=E_{\textrm{nuc}} + \sum_{p,q=1}^N k_{pq}  \hat{E}_{pq}
    + \frac{1}{2} \sum_{t=1}^{N_{\textrm{DF}}} \sum_{k,l=1}^N V^t_{kl} \hat{G}_t^{\dag}  \hat{E}_{kk} \hat{E}_{ll} \hat{G}_t.
\end{eqnarray}
We can now apply a
fermion-to-qubit mapping based on the Jordan-Wigner transformation \cite{jordan1928paul}. Since within this framework $\hat{E}_{kk}=\hat{I}-\frac{1}{2}(\hat{Z}_k+\hat{Z}_{\bar{k}})$ and
\begin{eqnarray}
    \hat{E}_{kk} \hat{E}_{ll} = -\hat{I}+ \hat{E}_{kk} + \hat{E}_{ll} + \frac{1}{4}
    (\hat{Z}_k+\hat{Z}_{\bar{k}}) (\hat{Z}_l+\hat{Z}_{\bar{l}}),
\end{eqnarray}
the Hamiltonian can be finally expressed as:
\begin{eqnarray}
    \hat{H} = \!\!\!\!\!\!\! & &\mathcal{E} - \frac{1}{2} \sum_{k=1}^N f^{\text{\o}}_k \hat{G}_{\text{\o}}^{\dag} (\hat{Z}_k+\hat{Z}_{\bar{k}})  \hat{G}_{\text{\o}} \nonumber \\
   & & + \frac{1}{8} \sum_{t=1}^{N_{\textrm{DF}}} \sum_{k,l=1}^N V^t_{kl} \hat{G}_t^{\dag} \left( \hat{Z}_k \hat{Z}_l -\delta_{kl} + \hat{Z}_k \hat{Z}_{\bar{l}} + \hat{Z}_{\bar{k}} \hat{Z}_l + 
     \hat{Z}_{\bar{k}} \hat{Z}_{\bar{l}} -\delta_{\bar{k}\bar{l}} \right) \hat{G}_t. \label{finalDFH}
\end{eqnarray}
In this equation, the one-electron tensor has been redefined as 
$f_{qp}=k_{qp}+\sum_{r=1}^N g_{pqrr}$ and a ``single factorized'' (eigenvalue) decomposition has been applied to
obtain
\begin{eqnarray}\label{mainsf}
    f_{pq}= \sum_{k=1}^N U^{\text{\o}}_{pk} f^{\text{\o}}_k U^{\text{\o}}_{qk}.
\end{eqnarray}
The constant term $\mathcal{E}$ contains $E_{\textrm{nuc}}$ and additional constant terms originating from the
one and two-body operators.

The block encoding of the double-factorized Hamiltonian can be obtained by 
straightforward application of the linear combination of unitaries (LCU) approach to the Hamiltonian in the form of
Eq.~\ref{finalDFH} \cite{childs2012hamiltonian,low2019hamiltonian}. In this case, the 1-norm takes the following value:
\begin{eqnarray}
    \lambda^{\textrm{LCU}}_{\textrm{DF}} = \sum_{k=1}^N |f^{\text{\o}}_k| + \frac{1}{2} \sum_{t=1}^{N_{\textrm{DF}}} \sum_{k,l=1}^N |V^t_{kl}| - 
    \frac{1}{4} \sum_{t=1}^{N_{\textrm{DF}}} \sum_{k=1}^N |V^t_{kk}|. \label{LCUnorm}
\end{eqnarray}

An alternative method to block encode the double-factorized Hamiltonian has been introduced
by von Burg \emph{et al.}~\cite{von2021quantum}. To introduce this
approach we assume that the $\mathbf{V}^t$ tensors have rank
one and are positive definite for every value of $t$. As discussed 
in the next Section, not all the approaches 
to build the double factorization satisfy these properties and
this has important repercussions on the efficiency of a specific method.
A positive-definite rank-one $\mathbf{V}^t$ 
can always be decomposed as
\begin{eqnarray}\label{decomposedZ}
    V^t_{kl} = W_{k}^t W_{l}^t; 
\end{eqnarray}
by replacing this factorization in Eq.~\ref{intermediateSQ} and reapplying the 
Jordan-Wigner transformation of the Hamiltonian takes the form
\begin{eqnarray}
    \hat{H} = \!\!\!\!\!\!\! & &\mathcal{E} - \frac{1}{2} \sum_{k=1}^{N} f^{\text{\o}}_k \hat{G}_{\text{\o}}^{\dag} (\hat{Z}_k+\hat{Z}_{\bar{k}})  \hat{G}_{\text{\o}} \nonumber \\
   & & + \frac{1}{8} \sum_{t=1}^{N_{\textrm{DF}}}   \hat{G}_t^{\dag} \left( \sum_{k=1}^{\Xi^{(t)}} W^t_{k} (\hat{Z}_k + \hat{Z}_{\bar{k}})  \right)^2 \hat{G}_t, \label{vonburgH}
\end{eqnarray}
where the sums over $k$ have been truncated at $\Xi^{(t)}\leq N$ by eliminating the elements of the $\mathbf{W}^t$ tensors below a certain threshold $\delta_{\textrm{DF}}$; $\Xi$, which denotes the average of the $\Xi^{(t)}$ values, has been used in the Introduction 
to discuss the computational complexity of double factorization (see Table \ref{tab:qubitizationcomplexity}). For this formulation of the electronic Hamiltonian we have $\Gamma=\mathcal{O}(N^2\Xi)$,
since in the two-body term we have a sum over $N_{\textrm{DF}}$ (which is itself $\mathcal{O}(N)$),
a sum over $\Xi$, and $N$ additional degrees of freedom in the basis rotations.
Within this formulation, the 1-norm of the Hamiltonian is given by
\begin{eqnarray}
    \lambda^{\textrm{Burg}}_{\textrm{DF}}=\sum_{k=1}^{N} |f^{\text{\o}}_k| + \frac{1}{4} \sum_{t=1}^{N_{\textrm{DF}}} \left( \sum_{k=1}^{\Xi^{(t)}} |W^t_{k}| \right)^2. \label{burgnorm}
\end{eqnarray}
This approach has two main advantages: (1) The Hamiltonian in Eq.~\ref{vonburgH} can
be efficiently implemented using qubitization \cite{low2019hamiltonian}; (2) the 1-norm in Eq.~\ref{burgnorm}
is typically significantly smaller with respect to the LCU 1-norm in Eq.~\ref{LCUnorm}.
As discussed in the following Sections of the paper, the implementation of
von Burg \emph{et al.}\ benefits significantly from the low rank of the
$\mathbf{V}^t$ tensor, and this will be an important feature included in 
our new SCDF methodology.

\subsection{Explicit and compressed double factorization}\label{sec:XDFandCDF}

In the previous Section, we introduced the general double factorization formalism and presented the benefits of this approach.
Here, we explain the main practical schemes that can be used to build the tensor factorization in Eq.~\ref{DFmain}. These approaches fall into two main 
categories: (1) explicit double factorization (XDF),
which builds the $\mathbf{V}^t$ and $\mathbf{U}^t$
tensors using a two-step eigenvalue or Cholesky decomposition \cite{motta2021low,kivlichan2018quantum,berry2019qubitization,huggins2021efficient}; (2) compressed double factorization (CDF) and its variants that build those tensors optimizing a cost function \cite{cohn2021quantum,oumarou2022accelerating}.  

Within the framework of XDF, the two-electron tensor is first decomposed
in terms of eigenvalues and eigenvectors:
\begin{eqnarray}\label{firstfact}
    (pq|rs)=\sum_{t=1}^{N_{\textrm{DF}}} V_{pq}^t \lambda_t V_{rs}^t = \sum_{t=1}^{N_{\textrm{DF}}} L_{pq}^t  L_{rs}^t
\end{eqnarray}
where we introduced the definition $L_{rs}^t \equiv \sqrt{\lambda_t} V_{rs}^t$.
A second factorization can then be obtained from the eigendecomposition of the $\mathbf{L}^t$ 
tensors:
\begin{eqnarray}\label{secondfact}
    L_{rs}^t=\sum_{k=1}^{\Xi^{(t)}} U_{rk}^t W^t_k U_{sk}^t, 
\end{eqnarray}
whose rank has been truncated to a $t$-dependent value $\Xi^{(t)}$, which is
at most equal to $N$.
The combination of these two equations provides a tensor decomposition in the form
of Eq.~\ref{DFmain} by defining
\begin{eqnarray}\label{ZfactXDF}
    V^t_{kl}\equiv W^t_k  W^t_l,
\end{eqnarray}
which is clearly consistent with the general definition in
Eq.~\ref{decomposedZ}.
According to Eq.~\ref{ZfactXDF} the $\mathbf{V}^t$ tensor has rank 1 and, as already mentioned in 
the previous Section, this is an important feature to implement efficiently
the double factorized Hamiltonian using the approach of von Burg \emph{et al.}~\cite{von2021quantum}. 
This XDF procedure is, in principle, exact if no truncation is applied, namely
 $N_{\textrm{DF}}=N^2$ and $\Xi^{(t)}=N$.
In practice, it is well known from widely used techniques such as density fitting or Cholesky decomposition\cite{dunlap1979some,werner2003fast,pedersen2009density,hohenstein2010density} that the rank of the two-electron integrals 
to achieve reasonable accuracy is much smaller than $N^2$ and in 
practice $N_{\textrm{DF}}$ scales as $\mathcal{O}(N)$.
Concerning the truncation of the second factorization, while the average rank 
$\Xi$ can be smaller than $N$ its complexity still behaves as $\mathcal{O}(N)$.

To further improve the scaling or at least the numerical complexity prefactor of the DF,
it is alternatively possible to build the Hamiltonian from $\mathbf{V}^t$ and
$\mathbf{U}^t$ tensors obtained by minimizing a suitable cost function. 
While $\mathbf{U}^t$ and $\mathbf{V}^t$ are constrained to still be orthogonal and symmetric,
respectively, this type of approach takes advantage of the full rank of these tensors.
This is the main idea at the base of 
compressed double factorization (CDF), which determines the DF tensors by minimizing the cost
function \cite{cohn2021quantum}
\begin{eqnarray}\label{costCDF}
    \mathcal{L}^{\textrm{CDF}}(\mathbf{U}^{t}, \mathbf{V}^t) = \frac{1}{2} \left | (pq|rs) - \sum_{t=1}^{N_{\textrm{DF}}} \sum_{k,l=1}^{N} U^{t}_{pk} U^{t}_{qk} V^t_{kl} U^{t}_{rl} U^t_{sl} \right |^2_{\mathcal{F}},
\end{eqnarray}
where $\mathcal{F}$ denotes the Frobenius norm. With respect to XDF, a smaller value of $N_{\textrm{DF}}$ can typically achieve the same level of accuracy in the final result.
Minimization of the cost
function can be achieved more efficiently by alternating the optimization of the $\mathbf{V}^t$ and $\mathbf{U}^t$ tensors.
The optimization with respect to $\mathbf{V}^t$
can be recast in the form of a linear system and solved with
standard linear algebra libraries. 
The orthogonality of the $\mathbf{U}^{t}$ tensors should be constrained
during the optimization. As explained in Appendix \ref{appendix:opt},
this problem can be reformulated in 
an unconstrained form
by introducing antisymmetric orbital rotation generators.
The derivatives of the cost function with respect to the components of the generators 
can be evaluated analytically and, since the minimization 
problem is non-linear in this case, a 
numerical unconstrained continuous optimizer such
as limited-memory Broyden–Fletcher–Goldfarb–Shanno
algorithm (L-BFGS) is used.

The CDF approach provides a more compact representation in terms of the $N_{\textrm{DF}}$ rank 
required for a given level of accuracy but
usually converges to tensors $\mathbf{V}^t$ with large components,
which in turn leads to
1-norms comparable or even larger with respect to XDF. 
For this purpose, a regularized compressed double factorization (RCDF) has been recently introduced 
that is based on the following cost function \cite{oumarou2022accelerating,goings2022reliably}:
\begin{eqnarray}
    \mathcal{L}^{\textrm{RCDF}}(\mathbf{U}^{t}, \mathbf{V}^t) = \frac{1}{2} \left | (pq|rs) - \sum_{t=1}^{N_{\textrm{DF}}} \sum_{k,l=1}^{N} U^{t}_{pk} U^{t}_{qk} V^t_{kl} U^{t}_{rl} U^t_{sl} \right |^2_{\mathcal{F}} + \sum_{t=1}^{N_{\textrm{DF}}} \sum_{k,l=1}^{N} \rho_{tkl} |V^t_{kl}|^{\gamma}, 
\end{eqnarray}
where the components of the tensor $\rho_{tkl}$ are usually fixed to a constant value and
$\gamma$ takes the values 1 or 2 for L1 and L2 regularization, respectively.
The last term in the cost function can be considered as a penalty function that prevents the elements of the $\mathbf{V}^t$ tensor from becoming too large and,
accordingly, limits the growth of the 1-norm. 
In Ref. \citenum{oumarou2022accelerating} all numerical applications were based on the
L2 regularization and this approach showed a sizeable decrease of the $\lambda^{\textrm{Burg}}_{\textrm{DF}}$ norm with respect to XDF and THC.

It is important to notice that in both the CDF and RCDF approaches, the $\mathbf{V}^t$ tensors are not necessarily positive definite, and their rank is unconstrained during the optimization. 
As discussed in Ref. ~\citenum{oumarou2022accelerating},
to generalize the approach of von Burg \emph{et al.}\
to the (R)CDF case, it is possible to introduce the 
factorization
\begin{eqnarray}\label{ZfactRCDF}
V^t_{kl} = \sum_{i=1}^N W_{ki}^t W_{li}^t.
\end{eqnarray}
However, this approach has two major disadvantages. First, the components of $\mathbf{W}^t$ are complex, requiring some modifications
of the original implementation of double factorization based on qubitization. Second, with respect to Eq.~\ref{decomposedZ}, the tensor decomposition in Eq.~\ref{ZfactRCDF} involves an additional sum over 
$N$ terms.
This implies that the amount of information $\Gamma$ defining the Hamiltonian
has complexity $\mathcal{O}(N^4)$, which is of the same order as the unfactorized Hamiltonian and
strongly affects the computational resource requirements. 
Using the RCDF tensor factorization provided in Ref.~\citenum{oumarou2022accelerating},
in Sec.~\ref{sec:P450}, we will show how Toffoli gate and logical qubit requirements
are affected in practice for the case of cytochrome P450.
Since our new SCDF method is based on a cost function
analogous to those used in CDF and RCDF, 
this issue is overcome by constraining the 
$\mathbf{V}^t$ tensor to be rank 1
during the optimization.

\subsection{Symmetry shift approach}\label{sec:symshift}

The main idea of symmetry shift involves replacing the
Hamiltonian $\hat{H}$ with $\hat{H}-\hat{S}(\mathbf{a})$,
where $\hat{S}(\mathbf{a})$ is a generic symmetry operator
that satisfies $[\hat{H},\hat{S}(\mathbf{a})]=0$ \cite{loaiza2022reducing,loaiza2023reducing}.
The set of parameters $\mathbf{a}$ is chosen to minimize
the 1-norm. Since the symmetry operator commutes
with the full electronic Hamiltonian, $\hat{H}- \hat{S}(\mathbf{a})$ has the same eigenstates of $\hat{H}$ and
can be directly used in QPE, taking advantage of the  
smaller 1-norm. The generic operator $\hat{S}(\mathbf{a})$ 
can be built as a function of one or multiple 
reciprocally commuting basic symmetry operators of which we know the eigenvalue for the ground state ahead of time.
The possible choices of these symmetry operators
include the number operator of electrons  $\hat{N}_e=\sum_{k=1}^{N} \hat{E}_{kk}$,
the z-projection of the spin,
the total spin, and the molecular 
point group symmetries. 
The original work of Loaiza and Izmaylov \cite{loaiza2022reducing, loaiza2023reducing} focused
on the $\hat{N}_e$ shift, which was shown to be   
effective to significantly reduce the 1-norm. 
In practice, this method 
is based on the 
shifted Hamiltonian
\begin{eqnarray}
    \hat{H}_S = \hat{H} - \hat{S}(\mathbf{a}) = \hat{H} - a_1 \hat{N}_e - a_2 \hat{N}_e^2, \label{LoIzshift}
\end{eqnarray}
where the two terms corresponding to $\hat{N}_e$ and $\hat{N}_e^2$ are intended to decrease the 1-norm of the one and two-body components of the Hamiltonian, respectively.
The strategy to optimize the parameters $\mathbf{a}$ is detailed in Ref.~\citenum{loaiza2022reducing}. For the one-body term of the Hamiltonian $\hat{H}_1$ we have
\begin{eqnarray}\label{symshift1}
    \hat{H}_1-a_1 \hat{N}_e &=& -\frac{1}{2} \sum_{k=1}^{N} (f_k^{\text{\o}}+2a_1) 
    \hat{G}_{\text{\o}}^{\dag} \hat{E}_{kk} \hat{G}_{\text{\o}} \nonumber \\
    &=& -\frac{1}{2} \sum_{k=1}^N (f_k^{\text{\o}}-a_1') 
    \hat{G}_{\text{\o}}^{\dag} \hat{E}_{kk} \hat{G}_{\text{\o}},
\end{eqnarray}
where we have used the fact that $\hat{N}_e$ commutes with the orbital rotation operator.
From the definition of the 1-norm in Eq.~\ref{burgnorm}, it is clear that the optimal $a_1'$ 
has to be chosen to minimize $\sum_{k=1}^N|f_k^{\text{\o}}-a_1'|$. The optimal
value can be simply obtained from
the median of the $f_k^{\text{\o}}$ coefficients.

Starting from Eq.~\ref{finalsq}, the symmetry shift for the two-body term can be written as
\begin{eqnarray}\label{symshift2}
    \hat{H}_2 - a_2 \hat{N}_e^2 &=& \frac{1}{2} \sum_{p,q,r,s=1}^{N} g_{pqrs} \hat{E}_{pq} \hat{E}_{rs} - a_2 \sum_{p,r=1}^N \hat{E}_{pp} \hat{E}_{rr} \nonumber \\
    &=& \frac{1}{2} \sum_{p,q,r,s=1}^N (g_{pqrs}-a_2'\delta_{pq}\delta_{rs}) \hat{E}_{pq} \hat{E}_{rs},  
\end{eqnarray}
where $a_2'=2a_2$.
Similarly to the one-body case, the optimal value can be found by minimizing $|g_{pqrs}-a_2'\delta_{pq}\delta_{rs}|$.
Once the two-electron integrals have been redefined, including the symmetry shift, the double-factorized two-body
Hamiltonian is obtained as in the regular XDF case, but the 1-norm is typically significantly reduced.

For several small molecules in the minimal STO-3G basis set, the XDF approach coupled with symmetry shift was shown to be a very
promising method both in terms of the 1-norm values and overall scaling of the 1-norm as a function of the system size \cite{loaiza2022reducing}. 
Indeed, this approach outperformed many others, such as orbital optimization, anti-commuting Pauli product grouping, and greedy Cartan sub-algebra decomposition both with and without symmetry shift.

Our new methodology, which we will introduce in the next Section, is exclusively focused on the 
1-norm reduction of the two-body part of the Hamiltonian. The approach of Eq.~\ref{symshift1} without modifications will be
used for the one-body term. 

\subsection{Symmetry-compressed double factorization}\label{sec:SCDF}

In this Section, we introduce our new approach, which will be denoted as
symmetry-compressed double factorization (SCDF). This method
combines some of the advantages of RCDF \cite{oumarou2022accelerating} and symmetry shift \cite{loaiza2022reducing,loaiza2023reducing} to 
significantly decrease the 1-norms of the Hamiltonian, which results in lower Toffoli gate counts. 
As a first step to introduce the SCDF approach, we focus on the two-body part of the Hamiltonian and consider the identity
\begin{eqnarray}
\hat{H}_2-\frac{\alpha}{2}  \hat{N}_e^2 & = &  \frac{1}{2}  \sum_{p,q,r,s=1}^N g_{pqrs} \hat{E}_{pq} \hat{E}_{rs} - \frac{\alpha}{2} \hat{N}_e^2 \nonumber \\
  & = & \frac{1}{2}  \sum_{p,q,r,s=1}^N g_{pqrs} \hat{E}_{pq} \hat{E}_{rs} - \frac{\alpha}{2} \sum_{p,r=1}^N  \hat{E}_{pp} \hat{E}_{rr} \nonumber \\ 
  & = & \frac{1}{2} \sum_{p,q,r,s=1}^N g_{pqrs} \hat{E}_{pq} \hat{E}_{rs} - \frac{1}{2} \sum_{t=1}^{N_{\textrm{DF}}} \sum_{p,q,r,s=1}^N \alpha^t \delta_{pq} \delta_{rs} \hat{E}_{pq} \hat{E}_{rs} \nonumber \\
  & = & \frac{1}{2} \sum_{t=1}^{N_{\textrm{DF}}} \sum_{k,l=1}^N \sum_{p,q,r,s=1}^N  U^{t}_{pk} U^{t}_{qk} V^t_{kl} U^{t}_{rl} U^t_{sl} \hat{E}_{pq} \hat{E}_{rs}
  - \frac{1}{2} \sum_{t=1}^{N_{\textrm{DF}}} \sum_{k,l=1}^N \sum_{p,q,r,s=1}^N U^{t}_{pk} U^{t}_{qk} \alpha^t U^{t}_{rl} U^t_{sl} \hat{E}_{pq} \hat{E}_{rs} \nonumber \\
  & = & \frac{1}{2} \sum_{t=1}^{N_{\textrm{DF}}} \sum_{k,l=1}^N (V^t_{kl}-\alpha^t) \hat{G}_t^{\dag}  \hat{E}_{kk} \hat{E}_{ll} \hat{G}_t, \label{N2shift}
\end{eqnarray}
where the Kronecker deltas in the third line were resolved
using the orthogonality of the $\mathbf{U}^t$ tensors (see Eq.~\ref{ortho}). 
The coefficient in front of $\hat{N}_e^2$ has been decomposed 
as $\alpha=\sum_{t=1}^{N_{\textrm{DF}}} \alpha^t$, where, in the ideal case, the differences $|V^t_{kl}-\alpha^t|$ should be as small as possible to effectively decrease 
the 1-norm.
There are two main differences between the symmetry shift technique proposed by Loaiza \emph{et al.}\ (Eq.~\ref{symshift2}) and the approach that we are proposing in 
Eq.~\ref{N2shift}: in Eq.~\ref{symshift2}, the symmetry shift is applied before the double factorization, while in our approach is applied after; in Eq.~\ref{N2shift}, the 
``global'' symmetry shift is decomposed into different $t$-dependent contributions. 
We empirically observed from
numerical experiments that 
Eq.~\ref{N2shift} is less effective 
than Eq.~\ref{symshift2} in reducing
the 1-norm, when applied within the XDF 
framework.
However, the $\mathbf{V}^t$ tensors themselves can
be optimized to decrease the fluctuations of $|V^t_{kl}-\alpha^t|$ and
this is the main idea of the SCDF approach.
In practice, this can be achieved by minimizing the cost
function 
\begin{eqnarray}
    \mathcal{L}(\mathbf{U}^t, \mathbf{V}^t) = \frac{1}{2} \left | (pq|rs) - \sum_{t=1}^{N_{\textrm{DF}}} \sum_{k,l=1}^{N} U^{t}_{pk} U^{t}_{qk} V^t_{kl} U^{t}_{rl} U^t_{sl} \right |^2_{\mathcal{F}} + \rho \sum_{t=1}^{N_{\textrm{DF}}} \sum_{k,l=1}^{N} |V^t_{kl}-\alpha^t|, \label{costSCDFfullrank}
\end{eqnarray}
with respect to the $\mathbf{V}^t$ and $\mathbf{U}^t$
tensors, and the prefactors $\alpha^t$ of the symmetry shift. The tuning of the regularization coefficient $\rho$ determines the trade-off between the accuracy in the tensor decomposition 
of the two-electron integrals and the decrease of 
$|V^t_{kl}-\alpha^t|$.

Similarly to RCDF, the formulation of 
the cost function in Eq.~\ref{costSCDFfullrank} leads to full rank $\mathbf{V}^t$
tensors that increase the
number of terms in the double factorized 
Hamiltonian by a factor $N$ with respect to the XDF approach.
To constrain the rank of $\mathbf{V}^t$
to be equal to 1, we factorize the tensor  
as in Eq.~\ref{decomposedZ}
and modify the SCDF cost function to be
\begin{eqnarray}
    \mathcal{L}^{\textrm{SCDF}}(\mathbf{U}^t, \mathbf{W}^t) = \frac{1}{2} \left | (pq|rs) - \sum_{t=1}^{N_{\textrm{DF}}} \sum_{k,l=1}^{N} U^{t}_{pk} U^{t}_{qk} W_{k}^t W_{l}^t U^{t}_{rl} U^t_{sl} \right |^2_{\mathcal{F}} + \rho \sum_{t=1}^{N_{\textrm{DF}}} \sum_{k,l=1}^{N} |W_{k}^t W_{l}^t-\alpha^t|, \label{costfRXDF1norm2} 
\end{eqnarray}
which is now optimized with respect to $\mathbf{W}^t$ rather than $\mathbf{V}^t$. The downside is that the easy single-step update of $\mathbf{V}^t$ is now replaced with a higher order dependence on $\mathbf{W}^t$. Because 
of the constraint on the rank of $\mathbf{V}^t$, the cost function of SCDF has less variational
freedom compared to other CDF methods. However, the numerical applications of Sec.~\ref{results} 
show that minima with very small 1-norms can be achieved at the price of 
a large number of iterations in the optimization of the cost function.

It is important to notice that while $V^t_{kl}=W_{k}^t W_{l}^t$ is rank 1, its shifted
counterpart $W_{k}^t W_{l}^t-\alpha^t$ is rank 2, unless $\alpha^t$ is 0 (see Appendix \ref{appendix:SCDFH} for a detailed discussion). From a numerical standpoint, this is effectively equivalent to doubling $N_{\textrm{DF}}$, with potential negative consequences on quantum computing resources. 
In practice, only a limited number $N_{\alpha}$ of $\alpha^t$'s have an optimized value different from 0 and, accordingly, the cost of the implementation does not significantly change with respect to the basic XDF. In the
resource estimation presented in Sec.~\ref{results},
the influence of these additional terms is taken into account. In the same Section, we will provide additional quantitative details on the number of required $\alpha^t$'s and their influence on the 1-norm.

An additional important observation is that, while 
the $\mathbf{V}^{t}$ tensors are positive definite 
by construction, $\mathbf{V}^{t}-\alpha^t$ can
have negative eigenvalues. Accordingly, implementing the double-factorized Hamiltonian
based on Eq.~\ref{vonburgH} requires some modifications. This point is discussed in Appendix \ref{appendix:SCDFH}.

\section{Results}\label{results}

\subsection{Computational details}\label{sec:compdetails}

The SCDF numerical results for all the other systems considered in this work are based on a Python implementation that
uses the JAX library \cite{jax2018github}. A description of the optimization procedure 
and the parameters used are provided in Appendix \ref{appendix:opt}. For all the numerical applications considered in this work, a value
of $10^{-5}$ for the regularization coefficient $\rho$ was found to be a reliable option. Indeed, if for example $\rho$ is set to $10^{-4}$, the norm can be effectively optimized but, at least in certain cases, chemical accuracy for ground state energies is not achieved; if instead the value of $\rho$ is decreased to $10^{-6}$, the minimization of the cost function is slow and this option should be considered only if a high level of accuracy is required.

The Toffoli gate and logical qubit requirements for the quantum implementation of XDF
and SCDF are estimated using the OpenFermion library \cite{mcclean2020openfermion}. For the FeMoco molecule and all the hydrogen chains independently of the size, 10 bits for state preparation and 16 bits for rotations were used \cite{lee2021even};
for cytochrome P450 we considered 10 bits for state preparation and 20 bits for rotations \cite{goings2022reliably}.
As discussed in Appendix \ref{appendix:SCDFH}, the quantum implementation of SCDF 
is analogous to XDF, and the same approach for resource estimation can be used. 

It is important to consider that OpenFermion's resource estimation for double factorized Hamiltonians is based on a QROM
algorithm that optimizes the number of Toffoli gates at the expense of a higher number of logical qubits \cite{low2018trading,berry2019qubitization,lee2021even}. This subroutine 
is used in different steps of the quantum implementation and tend to dominate the estimate of the total computational cost (this is especially the case for the rotation angle lookup).
By using auxiliary ancillae the data lookup
can be implemented using $L/k+b(k-1)$ Toffoli gates and $b(k-1)+\log(L/k)$ ancillae, where $L$ is the number of entries to load and $b$ the number of bits used to represent each entry. The parameter $k$,
that must be a power of 2, controls the trade-off between 
the number of the required Toffoli gates and of logical qubits.
If $k=1$, the conventional implementation is recovered, requiring
$L$ Toffoli gates and log(L) auxiliary qubits. 
To minimize the Toffoli gate count the $k$ is chosen as close as 
possible to the optimal value $\sqrt{L/b}$; 
with this choice both the Toffoli gate and logical qubit requirements behave as $\mathcal{O}(\sqrt{Lb})$.
The use of QROM for the rotation angle lookup tends to dominate the computational cost and overall scaling.
The specific cost for this task is $(N_{\textrm{DF}}\Xi)/k_r+N \beta(k_r-1)$ for the Toffoli gates and $N\beta(k_r-1)+\log(N_{\textrm{DF}}\Xi/k_r)$ for the auxiliary logical qubits, where $\beta$ is the number of bits used to represent each single angle and $k_r$ denotes the specific $k$ parameter used for this task. In this case the number of Toffoli gates is optimized by $k_r=\sqrt{\frac{N_{\textrm{DF}}\Xi}{N\beta}}$, which leads 
to a $\mathcal{O}(\sqrt{N_{\textrm{DF}}\Xi N})$ cost for both 
the number of required logical qubits and Toffoli gates.
By considering that $N_{\textrm{DF}}$ grows itself as $\mathcal{O}(N)$, this discussion explains the computational complexity of double factorization reported in Table \ref{tab:qubitizationcomplexity}. Since $N_{\textrm{DF}} \Xi N$  
corresponds to the amount of information $\Gamma$ contained in the Hamiltonian, the use of this approach to trade off Toffoli gates for logical qubits explains the $\mathcal{O}(\sqrt{\Gamma})$ complexity discussed in the Introduction.
While the choice of the optimal value of $k_r$ is crucial
to reduce the Toffoli gate requirements and scaling, depending on the specific system and the number of logical qubits available it may be of interest to find a different space-time balance. 
To this purpose, for the FeMoco and P450 active space models considered below we also present results for 
``suboptimal'' values of the $k_r$ parameter, that sizably decrease the number of required logical qubits while still providing a low number of Toffoli gates.

The rank of the first factorization $N_{\textrm{DF}}$ is chosen
to be a multiple of the number of orbitals N, from a minimum of 4$N$ to a maximum of 6$N$. The rank of the second factorization 
$\Xi$ is determined by eliminating the components of the
$\mathbf{W}^t$ tensor below the threshold $\delta_{\textrm{DF}}=10^{-4}$; this choice ensures a significant reduction of the terms in the Hamiltonian while preserving
a high level of accuracy. In applying the symmetry shift, only the
components $\alpha^t$ above the threshold $\delta_{\alpha}=10^{-3}$ are included.

\subsection{Active space model of the FeMoco molecule}\label{sec:FeMoCo}

We begin by applying our approach to simulate the ground state of an active space model of the FeMoco active site of nitrogenase, which plays a crucial role in understanding the mechanism of biological nitrogen fixation \cite{beinert1997iron}.
This system was identified as a potential killer application of quantum
computing because of its strong static correlation, which is challenging to
simulate with classical techniques~\cite{reiher2017elucidating}. The original active space studied 
by Reiher \emph{et al.}~\cite{reiher2017elucidating} was later improved by Li \emph{et al.}~\cite{li2019electronic}.
For the purpose of demonstrating the efficiency of the SCDF approach, we focus here on 
the Reiher Hamiltonian.
The ground state quantum calculation for the corresponding active space involves 54 electrons with $N=54$ spatial orbitals, amounting to 108 qubits or spin orbitals.

\begin{table*}[tb!]
\centering
\begin{tabular}{c|c|c|c|c|c|c}
\hline \hline
\multirow{2}{*}{Approach} & \multirow{2}{*}{$N_{\textrm{DF}}$ or $N_{\textrm{THC}}$} & \multirow{2}{*}{$\Xi$}    & CCSD(T)   & \multirow{2}{*}{$\lambda$ (Ha)}  &  Toffoli & Logical  \\ 
 &   &  & error  (mHa) &   & gates  & qubits  \\
\hline \hline
XDF & $4N$ & 54 & 0.24  & 293.9 & $9.6\times10^9$ & 3,722  \\
XDF & $5N$ & 54 & 0.28  & 295.3 & $1.0\times10^{10}$ & 3,724  \\ 
XDF & $6N$ & 54 & 0.12 & 296.0 & $1.1\times10^{10}$ & 3,724  \\ 
\hline 
XDF+sym. shift & $4N$ & 54 & 0.25  & 182.9 & $6.0\times10^9$ & 3,722  \\
XDF+sym. shift & $5N$ & 54 & 0.28  & 184.3 & $6.5\times10^{9}$ & 3,724  \\ 
XDF+sym. shift & $6N$ & 54 & 0.13 & 185.0 & $7.1\times10^{9}$ & 3,724  \\ 
\hline 
THC$^{a}$ & $350\approx 6.5N$ & / & -0.29 & 306.3 & $5.3\times 10^9$  & 2,142  \\ 
\hline 
SCDF &  $4N$ & 39 & 0.60  & 79.9 & $2.4\times10^9$ & 3,719  \\
SCDF &  $5N$ & 35 & 0.32  & 78.0 & $\mathbf{2.4\times10^{9}}$ & 3,722  \\ 
SCDF &  $6N$ & 31 & 0.36 & 77.9 & $2.5\times10^{9}$ & 3,722  \\ 
\hline
SCDF ($k_r=2$) &  $5N$ & 35 & 0.32  & 78.0 & $2.6\times10^{9}$ & \textbf{1,994}  \\
\hline \hline
\multicolumn{6}{l}{\small $^a$ Results from Ref. \citenum{lee2021even}.} \\
\end{tabular}
\caption{Resource estimates and correlation energy errors for the active space model of the
FeMoco molecule. Different approaches based on DF and THC are compared for different values 
of the rank used in the factorization. The best-performing methods in terms of Toffoli gates and
logical qubits are highlighted in bold (while SCDF with $N_{\textrm{DF}}=4N$ provides a slightly smaller Toffoli gate count than $N_{\textrm{DF}}=5N$, the result with the highest accuracy
in the correlation energy is considered). If not explicitly indicated, the SCDF results were obtained with the optimal value $k_r=4$.}
\label{tab:results_femoco}
\end{table*}

All of the methods compared in Table \ref{tab:results_femoco} involve truncations of the tensor ranks that control the trade-off between accuracy and computational efficiency. For the methods based on DF, the behavior with respect to the truncation of the first factorization is assessed
comparing the three values $N_{\textrm{DF}}=4N, 5N$, and $6N$. 
The average rank of the second factorization
$\Xi$ is also shown in Table \ref{tab:results_femoco}. Interestingly, 
for XDF without or with symmetry shift, the application of the threshold $\delta_{\textrm{DF}}=10^{-4}$
to eliminate the components of the $\mathbf{W}^t$ tensors
has minimal effects on the total number of terms in the Hamiltonian and, accordingly,
on $\Xi$. This is different for SCDF, where $\Xi$ decreases when increasing $N_{\textrm{DF}}$,
leading to a total number of terms in the Hamiltonian that grows slowly with $N_{\textrm{DF}}$ (this
implies also a rather steady number of Toffoli gates).
This behavior is likely to be related to the characteristics of the SCDF cost function (Eq.~\ref{costfRXDF1norm2}), where large components of the $\mathbf{W}^t$ tensor are penalized.
It is worth mentioning that in Ref.~\citenum{lee2021even} 
a different procedure was used to truncate the XDF tensor decomposition. All the $N^2$ terms
were initially maintained in the first factorization and the pruning 
was exclusively performed for the second factorization 
removing the $j$th components that satisfy $ \left ( \sum_{k=1}^N |W_k^{t}| \right ) |W_j^{t}|<\delta_{\textrm{DF}}'$ ($\delta_{\textrm{DF}}'$ serves the same purpose of $\delta_{\textrm{DF}}$, but they are not
strictly equivalent). Within this procedure, the truncation of the second factorization effectively decreases
also the rank of the first factorization $N_{\textrm{DF}}$.
The resource estimation for FeMoco in Ref.~\citenum{lee2021even} was performed 
choosing $\delta_{\textrm{DF}}'=0.00125$ which leads to $N_{\textrm{DF}}=360$ and $\Xi=36$.
At first sight, the behavior of $\Xi$ could seem radically different with respect
to our results reported in Table \ref{tab:results_femoco}. 
In practice, for the $\mathbf{W}^t$ tensors with the largest contribution to the
tensor factorization, the two procedures provide similar values of the $\Xi^{(t)}$ rank.
For $\mathbf{W}^t$'s of decreasing importance the procedure of Ref.~\citenum{lee2021even} 
tends to keep more tensors, even with small values of $\Xi^{(t)}$; 
this explains the larger number
of $N_{\textrm{DF}}$ and the smaller average rank $\Xi$
form in Ref.~\citenum{lee2021even}.
In practice, for FeMoco our truncation scheme provides slightly more accurate ground state energies and similar total numbers of terms in the Hamiltonian and resource estimations.

The errors in the ground state energy and the resource estimations
for the different approaches considered here are presented in Table \ref{tab:results_femoco}.
A reliable tensor factorization of the Hamiltonian should preserve a high level of accuracy and,  following Ref.~\citenum{lee2021even}, we consider the correlation energy of the coupled cluster
with singles, doubles, and perturbative triples (CCSD(T))
as an error metric. As shown in the fourth column of Table \ref{tab:results_femoco} the error 
of all the different approaches  
is well below the chemical accuracy threshold of 1.6~mHa for all the values of $N_{\textrm{DF}}$. 
The resources required by XDF agree with previous findings in the literature \cite{von2021quantum,lee2021even}.
The application of the symmetry shift significantly decreases the 1-norm 
to about 62\% of the initial XDF value, and this is reflected in 
a very similar way in the number of required Toffoli gates.
However, the symmetry-shifted XDF is still not competitive with the
THC approach for both the numbers of required Toffoli gates and logical qubits.
For the FeMoco model, the new SCDF approach provides a significant additional reduction of the 1-norm, with values amounting to about one quarter of those
obtained with the XDF and THC methods. Because of this significant
1-norm decrease, the SCDF approach
requires less than half the number of Toffoli gates (and, accordingly, runtime) compared to the state-of-the-art
THC method. Since the number of logical qubits weakly depends on the 1-norm ($\log\lambda_{\textrm{DF}}$ dependence), the SCDF and all the DF-based approaches tend to be equivalent with respect to the qubit requirements. 
However, if for SCDF the $k_r$ parameter is decreased from its optimal value of 4 to 2, 
the number of Toffoli gates increases only slightly, while the logical qubit count  
becomes the smallest among all methods listed in Table \ref{tab:results_femoco}.
The same procedure could be applied to the other approaches to further decrease their logical qubit requirements  but this would further increase their Toffoli gate count.
According to these observations,  
for the FeMoco active space model, the SCDF method can achieve a significant speed-up with respect to all the
other approaches, and also provide an excellent balance between the number of Toffoli gates and logical qubits.

As discussed in Sec.~\ref{sec:SCDF}, only a limited number of
symmetry shift prefactors $\alpha^t$ actually contributes to the 1-norm reduction.
Using the $\delta_{\alpha}=10^{-3}$ threshold for FeMoco only 5 $\alpha^t$'s
are retained for $N_{\textrm{DF}}=4N=216$, 
13 for $N_{\textrm{DF}}=5N=270$, and 
25 for $N_{\textrm{DF}}=6N=324$; while 
this number grows with $N_{\textrm{DF}}$, it remains limited 
and, as shown in Table \ref{tab:results_femoco}, does not have
a significant impact on the computational cost.
This truncation has a negligible effect on the 1-norm (differences of the order of $10^{-8}$ Ha). Still, the
symmetry shift plays a fundamental role in the 1-norm reduction:
for example, if all the symmetry shift prefactors are set to 0
for $N_{\textrm{DF}}=5N$, the 1-norm increases to 145.8 Ha.

As a concluding note of this Section, 
it is interesting to explore the effects of the thresholds $\delta_{\alpha}$ and $\delta_{\textrm{DF}}$ on the accuracy and computational
requirements of SCDF. If for $N_{\textrm{DF}}=5N$ these two thresholds are
set to 0 and all the terms of the factorization are included, the 1-norm and the correlation 
energy error are not affected (the error in the correlation
energy actually slightly increases to 0.33~mHa). Instead, the computational requirements sizeably increase to $3.8\times 10^{9}$
Toffoli gates and $7,180$ logical qubits.
This finding demonstrates the importance of the sparsity in the SCDF
tensor factorization, beyond the beneficial effects of the 1-norm reduction.

\subsection{Active space model of cytochrome P450}\label{sec:P450}

To further assess the efficiency and accuracy of the SCDF method,
we consider here the cytochrome P450, which has been proposed 
as a benchmark system for fault-tolerant quantum algorithms in
Ref.~\citenum{goings2022reliably}. The $(34\uparrow\!+29\downarrow \text{e},58\text{o})$ active space model of the
Cpd I species was chosen as an example. 
Resource estimates for the ground state calculation of the
P450 model are reported in Table \ref{tab:results_p450}.
The behavior of the different methods is similar
to what we already discussed for the FeMoco case. In particular, our new SCDF method  
decreases to less than one half the Toffoli gate requirements 
of THC, with an equivalent speed-up expected for the
the runtime of the quantum algorithm.
It is also important to mention that for this application,
the accuracy of THC is less systematic. As shown in the
Supplementary Information of Ref.~\citenum{goings2022reliably},
the CCSD(T) error tends to oscillate as a function of the THC
rank. For example, if the rank is increased to 380 (to be compared to
320, the value chosen as optimal in Ref.~\citenum{goings2022reliably}), the CCSD(T) error increases in absolute value to
-0.83~mHa. The correlation energy error of SCDF and other DF-based approaches is instead significantly smaller and less dependent on
$N_{\textrm{DF}}$.
Compared to other DF variants, 
SCDF also decreases the requirements in terms of 
logical qubits. Since the qubit requirements have only a weak dependence on the 1-norm, this reduction is mainly due to the decreased number of terms in the Hamiltonian due to the lower $\Xi$ rank of SCDF. Similarly to the FeMoco active space model, changing the $k_r$ parameter from its optimal value of 2 to 1, the number of logical qubits can be further decreased at the price of a higher
number of Toffoli gates. In this case, for $k_r=1$
the number of logical qubits required by SCDF lies between the two
THC results reported in Table \ref{tab:results_p450}.

For the P450 application with $N_{\textrm{DF}}=5N=290$, only
7 non-zero $\alpha^t$'s are included using the $\delta_{\alpha}=10^{-3}$
threshold. This truncation changes the 1-norm only by 0.3 Ha,
corresponding to 0.3\%. Without the symmetry shift contribution, 
the 1-norm of the SCDF approach would increase from 111.3 to 216.5.
Similarly to the case of the FeMoco molecule, if the 
 $\delta_{\alpha}$ and $\delta_{\textrm{DF}}$ thresholds are set to zero, the
1-norm and correlation energy error are minimally affected but the resource requirements sizeably increase to 
$6.7 \times 10^9$ Toffoli gates and
$4,924$ logical qubits.

Since the RCDF tensor decomposition for the same P450 active space model were provided with Ref.~\citenum{oumarou2022accelerating},
in Table \ref{tab:results_p450} we also include 
 the resource estimation for this methodology.
To obtain the Toffoli gate and logical qubit requirements we assume that the implementation of the complex components of the $\mathbf{W}^t$ tensors 
(see discussion about Eq.~\ref{ZfactRCDF}) do not involve any overhead costs with respect to the real case. 
As discussed in Sec.~\ref{sec:XDFandCDF}, the RCDF 
factorization involves a factor $N$ more 
terms in the Hamiltonian as
compared to XDF and, similarly, SCDF. In order to decrease the number of terms
we used also in this case the same $\delta_{\textrm{DF}}=10^{-4}$ threshold, but this procedure
is not effective in this case.
Despite the sizeable decrease of the 1-norm with respect to XDF, the cost of implementing $C_{\mathcal{W}[\hat{H}]}$
is much higher for RCDF, and this explains the high computational requirements shown in 
Table \ref{tab:results_p450}.

\begin{table*}[tb!]
\centering
\begin{tabular}{c|c|c|c|c|c|c}
\hline \hline
\multirow{2}{*}{Approach} & \multirow{2}{*}{$N_{\textrm{DF}}$ or $N_{\textrm{THC}}$}   & \multirow{2}{*}{$\Xi$}   & CCSD(T)   & \multirow{2}{*}{$\lambda$ (Ha)} &  Toffoli & Logical  \\ 
 &  &  & error  (mHa) &   & gates  & qubits  \\
\hline \hline
XDF & $4N$ & 57 & 0.12  & 472.2 & $1.9\times10^{10}$ & 4,922  \\
XDF & $5N$ & 57 & 0.069  & 472.7 & $2.1\times10^{10}$ & 4,926  \\ 
XDF & $6N$ & 57 & 0.060 & 472.9 & $2.2\times10^{10}$ & 4,925  \\ 
\hline 
XDF+sym. shift & $4N$ & 57 & 0.12  & 298.9 & $1.2\times10^{10}$ & 4,920  \\
XDF+sym. shift & $5N$ & 57 & 0.066  & 299.4 & $1.3\times10^{10}$ & 4,924  \\ 
XDF+sym. shift & $6N$ & 57 & 0.057 & 299.6 & $1.4\times10^{10}$ & 4,923  \\ 
\hline 
RCDF$^a$ & $100\approx 1.7 N$ & 58 & 0.019 & 284.1 & $4.6 \times 10^{10}$  & 18,856 \\
\hline 
THC$^b$ & $320\approx 5.5 N$ & / & 0.10 & 388.9 & $7.8\times 10^9$  & \textbf{1,434}  \\ 
THC$^b$ & $380\approx 6.6 N$ & / & -0.83 & 392.5 & $8.3\times 10^{9}$ & 2,158 \\
\hline 
SCDF & $4N$ & 33 & -0.10  & 112.3 & $3.9\times10^9$ & 2,590  \\
SCDF & $5N$ & 26 & 0.044  & 111.3  & $\mathbf{3.8\times10^{9}}$ & 2,596 \\ 
SCDF & $6N$ & 24 & 0.069 & 111.0 & $4.0\times10^{9}$ & 2,594  \\ 
\hline
SCDF ($k_r=1$) & $5N$ & 26 & 0.044  & 111.3  & $4.8\times10^{9}$ & 1,706 \\
\hline \hline
\multicolumn{6}{l}{\small $^a$ The resource estimation was obtained using the tensors provided with Ref. \citenum{oumarou2022accelerating}.} \\
\multicolumn{6}{l}{\small $^b$ Results from Ref. \citenum{goings2022reliably}.}
\end{tabular}
\caption{Resource estimates and correlation energy errors for the active space model of 
cytochrome P450. Different approaches based on DF and THC are compared for different values 
of the rank used in the factorization. The best performing methods in terms of Toffoli gates and
logical qubits are highlighted in bold. If not explicitly indicated, the SCDF results were obtained with the optimal value $k_r=2$.}
\label{tab:results_p450}
\end{table*}

\subsection{Hydrogen chains}\label{sec:hchains}

In this Section, we discuss the scaling of the SCDF method for 
systems of growing size by considering hydrogen chain models
with up to 80 atoms. To compare with previous results in the literature
we use the same interatomic distance (1.4 Bohr) of Ref.~\citenum{lee2021even}
and the STO-6G basis set. For this specific choice of the basis set,
the number of spatial orbitals $N$ is equal to the number of hydrogen atoms $N_H$. 
For all the different DF-based approaches,  
$N_{\textrm{DF}}$ is set to $4N$ and the $\delta_{\textrm{DF}}=10^{-4}$ threshold is used to truncate the components in the second factorization. 
Our results are compared with 
the THC results from Ref.~\citenum{lee2021even}, where the tensor decomposition rank was set to 
$7N$. 

First of all, for systems of growing size, it is of fundamental importance to establish 
the accuracy in the prediction of the ground-state energy. Table \ref{tab:ccerror}
shows the error in the CCSD(T) correlation energy per atom in Hartree.
The SCDF method is characterized by errors of the order of $10^{-7}$ Ha per atom
and a maximum total (absolute) deviation of $5.06 \times 10^{-5}$ for H$_{80}$.
The XDF method without and with symmetry shift achieves
an even higher level of accuracy. In Ref.~\citenum{lee2021even} the authors aimed at achieving an
accuracy in the energy per atom within 50–60 $\mu$Ha. 
This choice leads to an
energy error per atom for THC that is about one or two
orders of magnitude larger than what we report here for SCDF. Importantly, these errors present a rather
erratic behavior as a function of N$_H$ and, for H$_{60}$, the total deviation in the CCSD(T) correlation energy even achieves the value of $3.1 \times 10^{-3}$~Ha, which largely exceeds the chemical accuracy
threshold ($1.6\times 10^{-3}$~Ha). While this level of accuracy for THC might 
not be sufficient to obtain accurate energy differences or properties in the thermodynamic limit, we nevertheless consider this factorization for the purpose of comparing
the resource requirements of 
THC and DF-based methods.

\begin{table*}[tb!]
\centering
\begin{tabular}{c|a|a|a|a}
\hline \hline
$N_{H}$      & XDF  & XDF+sym. shift  & THC$^a$  &  SCDF  \\ 
\hline \hline
10 & $7.2\times10^{-10}$  & $-5.8\times10^{-10}$ & $4.4\times10^{-6}$ & $-2.1\times10^{-7}$ \\
20 & $8.1\times10^{-10}$  & $8.3\times10^{-10}$ & $1.7\times10^{-5}$ & $8.7\times10^{-7}$  \\ 
30 & $-4.6\times10^{-8}$  & $-3.4\times10^{-8}$ & $-2.9\times10^{-7}$ & $7.7\times10^{-7}$  \\ 
40 & $1.1\times10^{-7}$  & $-2.9\times10^{-9}$ & $1.2\times10^{-5}$ & $-2.8\times10^{-7}$  \\
50 & $5.5\times10^{-8}$  & $8.7\times10^{-9}$ & $5.4\times10^{-6}$ & $-2.0\times10^{-7}$  \\ 
60 & $-6.8\times10^{-9}$  & $-1.8\times10^{-8}$ & $5.1\times10^{-5}$ & $1.5\times10^{-7}$  \\ 
70 & $-6.1\times10^{-8}$  & $-2.1\times10^{-8}$ & $1.9\times10^{-5}$ & $-3.2\times10^{-7}$ \\
80 & $-1.1\times10^{-7}$  & $-3.63\times10^{-10}$ & $4.9\times10^{-6}$ & $-6.3\times10^{-7}$  \\
\hline \hline
\multicolumn{5}{l}{\small $^a$ Results from Ref. \citenum{lee2021even}.}
\end{tabular}
\caption{CCSD(T) correlation energy error per atom (in Ha) for hydrogen chains with up to 80 atoms. Four methods are compared: explicit double factorization (XDF), XDF with symmetry shift, tensor hypercontraction (THC), and  
symmetry-compressed double factorization (SCDF).
}
\label{tab:ccerror}
\end{table*}

\begin{figure}[tb!]
    \centering
    \includegraphics[width=0.7 \textwidth]{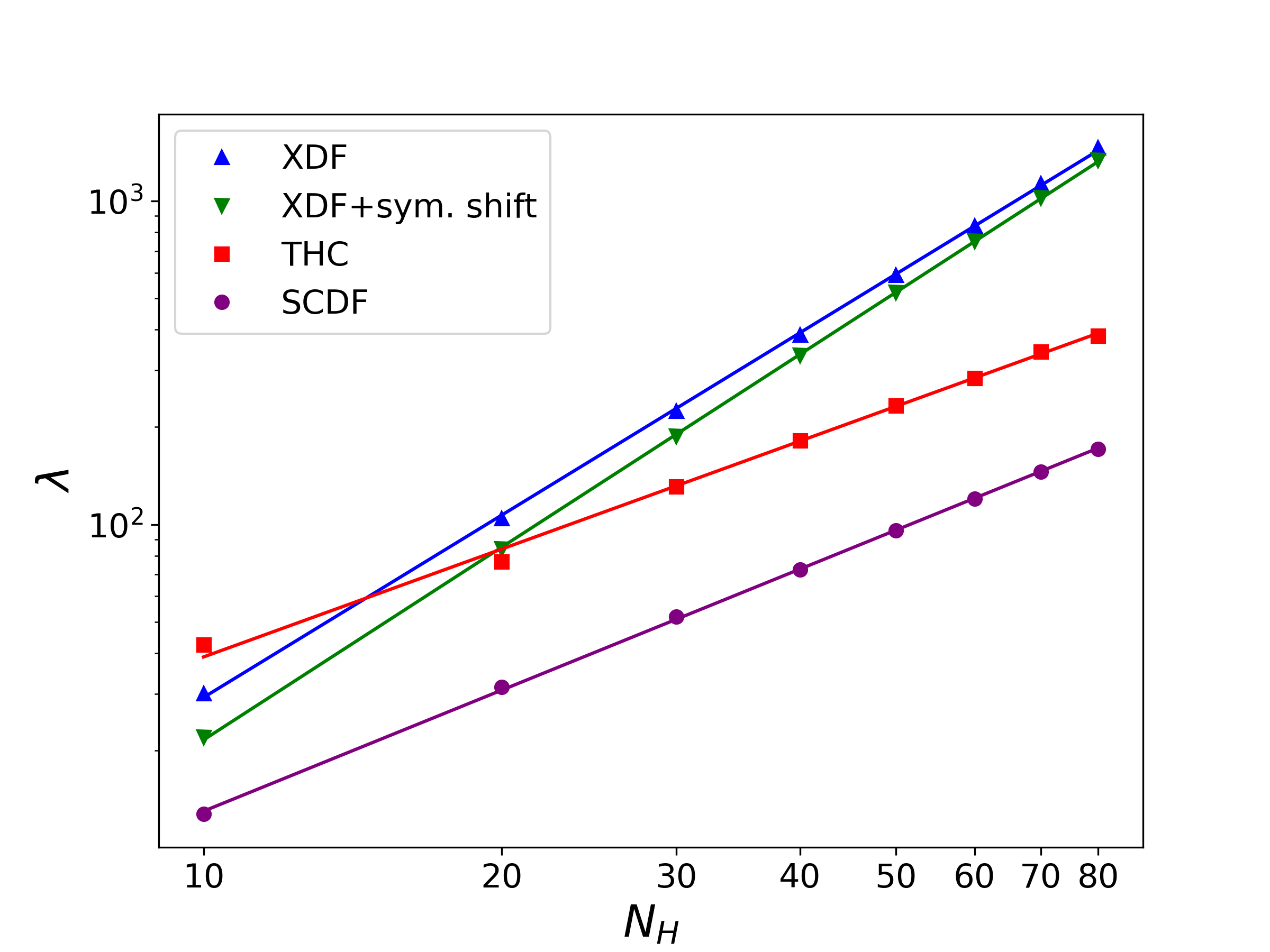}
    \caption{Scaling of the 1-norm $\lambda$ (in Ha) as a function of the number of hydrogen atoms $N_H$ for XDF, XDF with symmetry shift, THC, and SCDF.}
    \label{fig:onenormfit}
\end{figure}

\begin{figure}
  \begin{subfigure}{0.4\textwidth}
    \includegraphics[width=\linewidth]{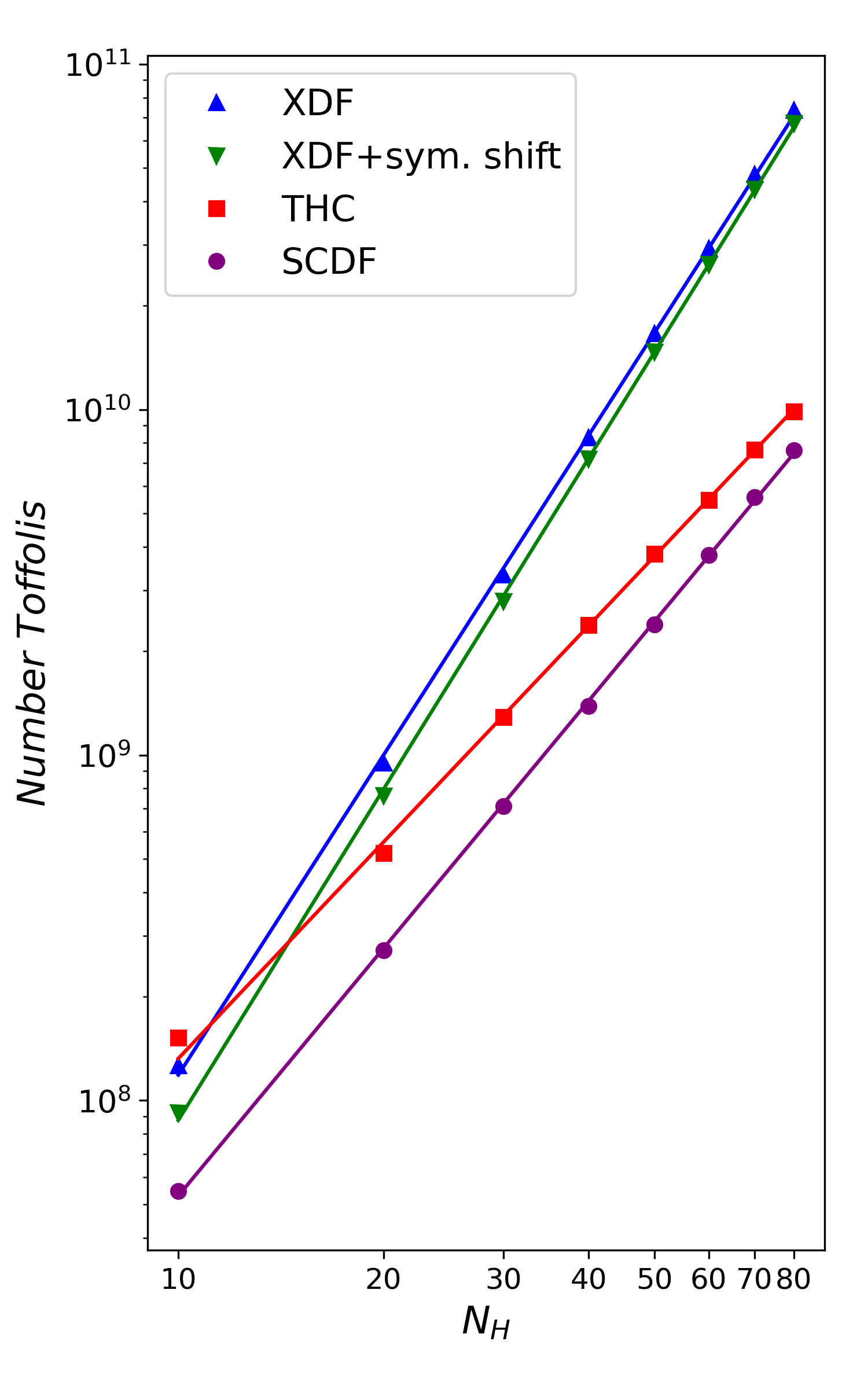}
  \end{subfigure}
  \begin{subfigure}{0.4\textwidth}
    \includegraphics[width=\linewidth]{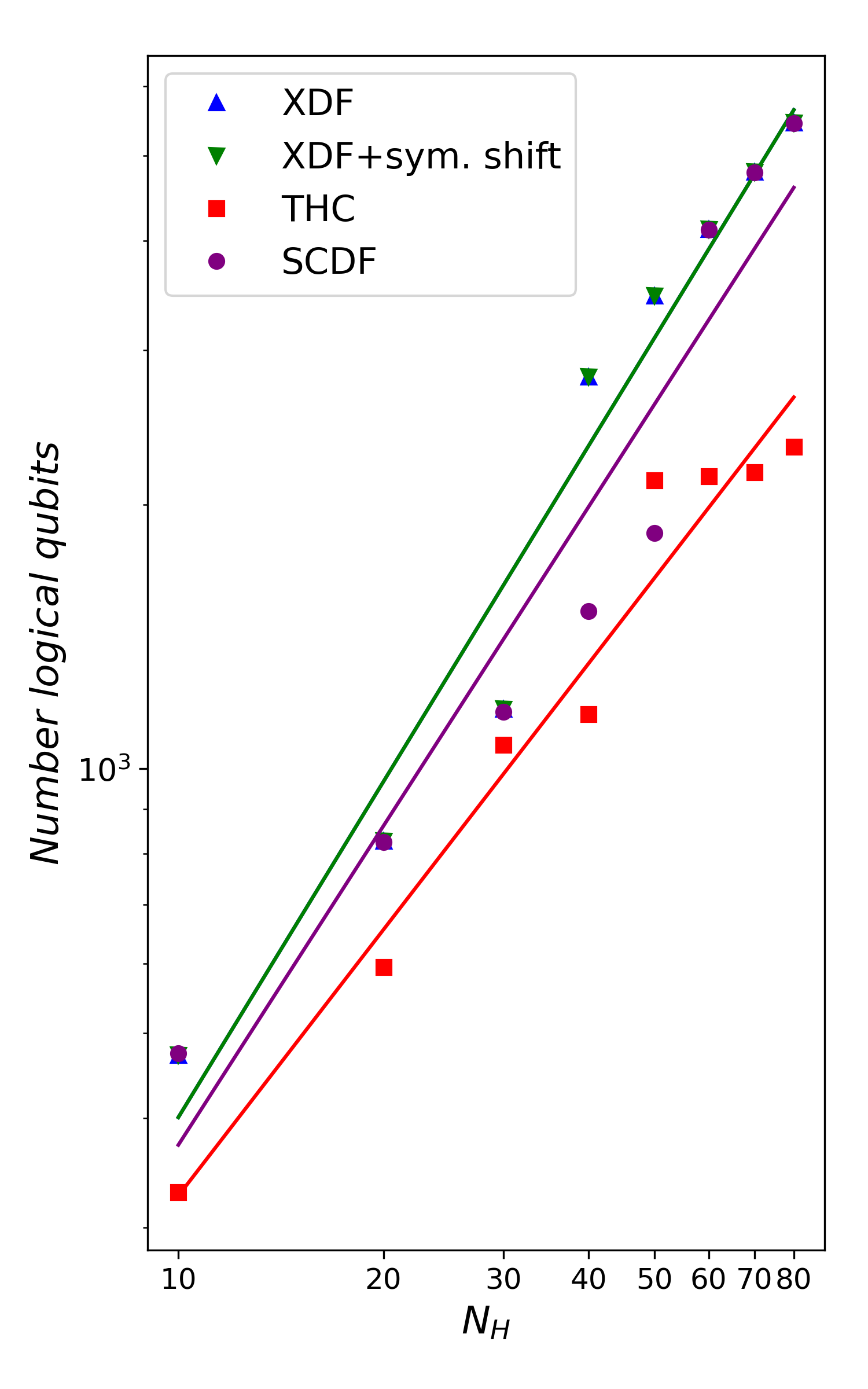}
  \end{subfigure}
  \caption{Left: Scaling of the number of Toffoli gates required
  by the XDF, XDF with symmetry shift, THC, and SCDF approaches as a function of the number of hydrogens $N_H$. Right: Scaling of the number of logical qubits.}
  \label{fig:toffqubitscaling}
\end{figure}

The 1-norm of the Hamiltonian 
plays a fundamental role in determining 
the number of QPE iterations and its scaling
has strong implications for the efficiency of
a specific method.
Fig.~\ref{fig:onenormfit} and Table \ref{tab:slopes} 
show the empirical behavior of the
1-norm as a function of N$_H$.
The 1-norms of the standard XDF approach and its symmetry-shifted counterpart do not significantly differ for the hydrogen chains and are characterized by 
a $\mathcal{O}(N^2)$ asymptotic complexity. This is decreased considerably
for SCDF, which has a similar scaling as THC 
and is expected to significantly decrease the 
computational requirements as compared to other DF-based methods. To this purpose, the left panel of Fig.~\ref{fig:toffqubitscaling} and Table \ref{tab:slopes} show
the dependence of the number of the Toffoli gates on the number of hydrogen atoms.
The SCDF approach requires significantly
less Toffoli gates than XDF (with or without symmetry shift)
and improves over their computational complexity by approximately a factor $N$, as expected by the 1-norm behavior. 
Within this range, SCDF also outperforms THC, but with 
a tendency to grow more rapidly.
Beyond the slightly different growth of $\lambda$, this should be expected as the theoretical scaling 
of the Toffoli gate requirements per iteration
for the DF-based methods behave as $\mathcal{O}(N\sqrt{\Xi})$ while 
for THC behave as $\mathcal{O}(N)$; although the hydrogen chains considered here are still too small to reproduce exactly these asymptotic behaviors, the slopes in Table \ref{tab:slopes} already reflect the trends correctly.

\begin{table*}[tb!]
\centering
\begin{tabular}{|c|a|a|a|a|a|a|}
\hline \hline
\multirow{2}{*}{Approach} & \multicolumn{2}{c|}{$\lambda$} & \multicolumn{2}{c|}{Number Toffolis} & \multicolumn{2}{c|}{Number logical qubits} \\
\hhline{|~|------|}
    & Slope & $R^2$ & Slope & $R^2$ & Slope & $R^2$  \\ 
\hline \hline
XDF & 1.87  & 0.9998 & 3.08  & 0.9997 & 1.27  & 0.9647 \\
XDF+sym. shift & 1.98  & 0.9999 & 3.18  & 0.9998 & 1.27  & 0.9647 \\ 
THC$^a$ & 1.11  &  0.9991 & 2.08  &  0.9998 & 1.01  &  0.9653  \\ 
SCDF & 1.24  &  0.9998 & 2.38  &  0.9998 & 1.21  &  0.9249 \\
\hline \hline
\multicolumn{5}{l}{\small $^a$ Results obtained using the data provided with Ref. \citenum{lee2021even}.}
\end{tabular}
\caption{Slopes of the linear fits on a logarithmic scale
of the 1-norm $\lambda$, the number of Toffoli gates, and the
number of logical qubits as a function of
the number of hydrogen atoms N$_H$.}
\label{tab:slopes}
\end{table*}

The behavior of the number of logical qubits as a function of the system size is less smooth and the fitting on a logarithmic scale in Table \ref{tab:slopes} and the right panel of Fig. \ref{fig:toffqubitscaling} has to be considered as only indicative of the overall trends. Since the number of logical
qubits has only a weak dependence on the 1-norm, in this case, the behaviors of the SCDF does not significantly differ with respect to XDF, and, as in the previous examples, THC has a lower
requirements than the DF methods.

\section{Conclusions}

In conclusion, we have introduced the symmetry-compressed double factorization approach
that couples the symmetry shift technique with regularized double factorization to decrease the 1-norm of the electronic Hamiltonian significantly. As the 1-norm determines the total number of iterations needed in QPE, its reduction
is important in decreasing the resource requirements and the runtime of fault-tolerant quantum simulations in chemistry.
The effectiveness of the SCDF method in reducing the 1-norm is demonstrated numerically with applications to different chemical systems, including active space models of the FeMoco molecule, cytochrome P450, and hydrogen chains up to 80 atoms. For these systems, the 1-norm values achieved by SCDF
are significantly lower than other methods, including XDF,
XDF with symmetry shift, RCDF, and THC.

Despite the fundamental role of the 1-norm, other factors should also be taken into account to assess the performance of a specific Hamiltonian factorization in the context
of qubitization-based QPE.

First, the cost of a single QPE iteration, which depends on the specific implementation and the number of terms in the Hamiltonian, can significantly impact the global computational cost.
To perform an unbiased comparison of methods, the Toffoli gate 
and logical qubit requirements were estimated for the chemical systems considered here. This shows that SCDF still outperforms all the
other methods regarding Toffoli gate counts. For example, for the FeMoco molecule and cytochrome P450,
SCDF requires fewer than 50\% of the Toffoli gates compared to THC, thus far recognized as the best-performing method in the literature.
However, the number of terms in the THC-factorized Hamiltonian
grows with lower complexity as compared to all other methods,
and this approach will tend to become the most efficient for
systems of growing size.
Concerning the logical qubit count, SCDF
inherits the same behavior as the other DF-based approaches,
and tends to require relatively large numbers of qubits.

Second, given that Hamiltonian factorizations frequently employ tensor truncations to enhance efficiency,
it is crucial to determine their impact on the ground-state total energy.
We considered CCSD(T) correlation energies as an accuracy metric consistent with prior literature.
Despite the potentially disruptive effect of the regularization, SCDF maintains a high level of accuracy comparable to other DF methods. The behavior of THC is
more problematic. For example, in the case of cytochrome P450
the error in the correlation energy exhibits an erratic behavior as a function of the THC rank, with a tendency to significantly increase after reaching very small values. The behavior of the correlation energy is also nonsystematic as a function of the size of the hydrogen chains, and, in at least one case, the error significantly exceeds the chemical accuracy threshold.

In this work, the significant 1-norm reduction provided by SCDF has been
exclusively exploited in the context of QPE.
However, different quantum algorithms could benefit from 
the 1-norm decrease. This is the case, for example, of
the stochastic compilation protocol known as qDRIFT \cite{campbell2019random}, which requires $O(\lambda^2)$ 
repetitions to approximate the time evolution.
The number of shots required to measure the expectation value of the Hamiltonian also grows with the
1-norm and, accordingly, 
also VQE and other near-term quantum algorithms could benefit from
its reduction. Establishing the accuracy and efficiency of 
SCDF, also for near-term quantum algorithms, will be the subject of
future work.
In the context of fault-tolerant quantum computing, the ideas
developed in this work may also apply within the THC framework.
While not straightforward, this extension could lead to
a highly efficient methodology with optimal scaling.

\section*{Acknowledgements}
We thank Nicholas Rubin for explaining the Openfermion resource estimation code and the details of THC, William Poll and Mark Steudtner for insightful discussions on DF and THC circuits, and Oumarou Oumarou and Christian Gogolin for valuable discussions on the RCDF method.

\appendix

\section{Optimization of the SCDF cost function}\label{appendix:opt}

The SCDF cost function in Eq.~\ref{costfRXDF1norm2} 
has to be minimized with respect to the components of 
the $\mathbf{W}^t$, $\mathbf{U}^t$, and $\alpha^t$ tensors.
As previously discussed for CDF \cite{cohn2021quantum},
the simultaneous optimization of the different tensors 
is inefficient and a nested approach with multiple steps should be preferred.
For the SCDF cost function the optimization
is performed through the following steps:
\begin{enumerate}
    \item Optimization of the $\mathbf{W}^t$ tensor using the unconstrained minimization L-BFGS algorithm;
    \item Update of the $\alpha^t$ values as median of the new $\mathbf{V}^t$;  
    \item Optimization of the $\mathbf{U}^t$ tensor using the L-BFGS algorithm.
    \item Repeat from step 1 until convergence is achieved.
\end{enumerate}

For the optimizations in steps 1 and 3 
we have developed two different implementations
based on analytical gradients and automatic differentiation
using the JAX library \cite{jax2018github}. While the former implementation is mainly
intended to provide a reference and to test the soundness of the numerics,  
the latter is significantly more efficient and is used in production runs. 
The convergence criterion in step 4 is naturally defined in terms of 
thresholds on gradients or on the decrease of the cost function
between subsequent iterations. However, since the 1-norm reduction
is crucial to enhance the QPE efficiency and its value is
found to decrease monotonically as a function of the
iteration count, the optimization procedure was stopped 
only when the 1-norm value was steadily decreasing well below
the 0.05 Ha threshold.

The analytical gradient of the SCDF cost function with respect to the components of
the $\mathbf{W}^t$ tensor is given by
\begin{eqnarray}
    \frac{\partial \mathcal{L}^{\textrm{SCDF}}}{\partial W_k^t} &=&
    -2 \sum_{p,q,r,s=1}^N \sum_{l=1}^N \Delta_{pqrs} U^t_{pk} U^t_{qk} W_l^t U^t_{rl} U^t_{sl}
    +2 \rho \sum_{l=1}^N \text{sign}(W_{k}^t W_{l}^t-\alpha^t) W_l^t 
\end{eqnarray}
where
\begin{eqnarray}
    \Delta_{pqrs} \equiv (pq|rs) - \sum_{t=1}^{N_{\textrm{DF}}} \sum_{k,l=1}^N U^{t}_{pk} U^{t}_{qk} W_{k}^t W_{l}^t U^{t}_{rl} U^t_{sl}.
\end{eqnarray}

The analytical gradient of the SCDF cost function with respect to the components of $\mathbf{U}^t$ is given by
\begin{eqnarray}
    \frac{\partial \mathcal{L}^{\textrm{SCDF}}}{\partial U_{pk}^t} &=&
    -4 \sum_{q,r,s=1}^N \sum_{l=1}^N \Delta_{pqrs}  U^t_{qk} W_l^t U^t_{rl} U^t_{sl}.
\end{eqnarray}
Since the regularization term does not depend on the the $\mathbf{U}^t$
tensors this derivative is the same as for the CDF and RCDF approaches.
The optimization of $\mathbf{U}^t$ is more complex than
in the $\mathbf{W}^t$ case since this
tensor has to be constrained to be orthogonal.
In practice this problem can be formulated as an unconstrained 
optimization by introducing the antisymmetric orbital rotation generators
matrices $\mathbf{X}^t$ to define $\mathbf{U}^t=\exp(\mathbf{X}^t)$.
The minimization of the cost function is then performed with respect to
the components of $\mathbf{X}^t$; the procedure is described in detail in
Ref. \citenum{cohn2021quantum}.

The minimization of the SCDF cost function is susceptible to the presence of local minima. Significantly lower values of the SCDF cost function and 1-norm can be found if very tight values ($10^{-12}$) are used for the tolerance of the stopping criterion of the L-BFGS algorithm. However, this approach requires a large number 
of iterations for each minimization of $\mathbf{W}^t$ and $\mathbf{U}^t$,
and the overall optimization is slow.

\section{Quantum implementation of the symmetry-compressed
double factorized Hamiltonian} \label{appendix:SCDFH}

The implementation of the (rank 1) double factorized Hamiltonian in the form of Eq.~\ref{vonburgH} has been discussed in details in Ref. \citenum{von2021quantum}. Specifically, this approach is valid for a positive definite/rank one $\mathbf{V}^t$ tensor, as in the 
XDF case.
In this appendix we show that the implementation of SCDF requires minimal modifications
with respect to the original work of von Burg \emph{et al.}.
Within the SCDF framework, Eq.~\ref{N2shift} redefines the two-body part
of the Hamiltonian by introducing a symmetry shift of each 
term in the sum over the index $t$, 
leading to the following form of Eq.~\ref{intermediateSQ}:
\begin{eqnarray}\label{intermediateSQshifted}
    \hat{H}=E_{\textrm{nuc}} + \sum_{p,q=1}^N k_{pq}  \hat{E}_{pq}
    + \frac{1}{2} \sum_{t=1}^{N_{\textrm{DF}}} \sum_{k,l=1}^N (V^t_{kl}-\alpha^t) \hat{G}_t^{\dag}  \hat{E}_{kk} \hat{E}_{ll} \hat{G}_t.
\end{eqnarray}
As shown by the numerical applications part in Sec. \ref{results},
the values of $\alpha^t$ are actually different from zero only for a small number $N_{\alpha}$ of indexes $t$.
In SCDF $\mathbf{V}^t=\mathbf{W}^t\otimes \mathbf{W}^t$ is rank 1 and the
element-wise constant shift $\alpha^t$ is equivalent to subtracting the
rank 1 tensor $-\alpha^t\mathbf{1}\otimes \mathbf{1}$ (here $\otimes$ 
denotes the outer product and $\mathbf{1}$ is a vector whose $N$ components are all 1). This implies that 
$\mathbf{V}^t-\alpha^t$ has rank two and, by applying
an eigenvalue decomposition, can be expressed as
\begin{eqnarray}
    \mathbf{V}^t-\alpha^t=\mathbf{P}^t \otimes \mathbf{P}^t - \mathbf{Q}^t \otimes \mathbf{Q}^t,
\end{eqnarray}
where $\mathbf{P}^t$ and $\mathbf{Q}^t$ are 1D tensors.
To keep the formalism consistent, in this appendix we will define 
$\mathbf{P}^t=\mathbf{W}^t$ and $\mathbf{Q}^t=\mathbf{0}$ for the $t$ indexes
corresponding to $\alpha^t=0$ (or, more precisely, corresponding to the $\alpha^t$'s smaller
than a given threshold $\delta_{\alpha}$). 
The application of the Jordan-Wigner transformation to the Hamiltonian in Eq.~\ref{intermediateSQshifted} leads to
\begin{eqnarray}
    \hat{H} = \!\!\!\!\!\!\! & &\mathcal{E} - \frac{1}{2} \sum_{k=1}^N f^{\text{\o}}_k G_{\text{\o}}^{\dag} (\hat{Z}_k+\hat{Z}_{\bar{k}})  G_{\text{\o}} \nonumber \\
   & & + \frac{1}{8} \sum_{t=1}^{N_{\textrm{DF}}}   \hat{G}_t^{\dag} \left( \sum_{k=1}^{\Xi^{(t)}} P^t_{k} (\hat{Z}_k + \hat{Z}_{\bar{k}})  \right)^2 \hat{G}_t
   - \frac{1}{8} \sideset{}{'}\sum_{t=1}^{N_{\textrm{DF}}}   \hat{G}_t^{\dag} \left( \sum_{k=1}^{\Theta^{(t)}} Q^t_{k} (\hat{Z}_k + \hat{Z}_{\bar{k}})  \right)^2 \hat{G}_t,
   \label{vonburgHshifted}
\end{eqnarray}
where the primed sum $\sideset{}{'}\sum$ indicates that only 
the $N_{\alpha}$ non-zero terms are actually included.
Similarly to Eq.~\ref{vonburgH}, a threshold $\delta_{\textrm{DF}}$ is applied 
to eliminate the small components of the $\mathbf{P}^t$ and $\mathbf{Q}^t$ tensors; this leads
to sums over $k$ that run up to values $\Xi^{(t)}$ and $\Theta^{(t)}$ that are less than or equal to $N$.
This equation is analogous to the 
formulation of von Burg \emph{et al.} in Eq.~\ref{vonburgH}, with only $N_{\alpha}$
additional terms with a negative sign. The implementation 
of terms with a negative sign
requires minimal modifications with respect to the implementation 
described in the supporting information (SI) of Ref. \citenum{von2021quantum}. 
The minus sign can be included when combining the two body terms
evaluated from qubitization. Using the same notation of the
SI of Ref. \citenum{von2021quantum}, this can be achieved by replacing
$\left |\overrightarrow{\Lambda_{SH}} \right \rangle^\dag$ with 
$\overline{\left |\overrightarrow{\Lambda_{SH}} \right \rangle}^{\ \dag}$ in the circuit in Eq. 78 therein. The notation with an 
additional line on top is used in Ref.~\citenum{von2021quantum} to distinguish Eqs.~7 and 8, which correspond to the prepare operator for block encoding 
without or with sign, respectively.  
The quantum circuits for $\left |\overrightarrow{\Lambda_{SH}} \right \rangle^\dag$ and $\overline{\left |\overrightarrow{\Lambda_{SH}} \right \rangle}^{\ \dag}$ are very similar and require the same number of Toffoli gates. Accordingly, the required resources 
are not affected by the
negative sign and the resource
estimation was carried out using the standard OpenFermion implementation \cite{mcclean2020openfermion}.

\bibliography{SCDF}

\end{document}